\pdfoutput=1
\documentclass[11pt,a4paper]{article}

\usepackage[T1]{fontenc}
\usepackage[utf8]{inputenc}
\usepackage{lmodern}
\usepackage[margin=2.4cm]{geometry}
\usepackage{graphicx}
\usepackage{booktabs}
\usepackage{array}
\usepackage{multirow}
\usepackage{amsmath}
\usepackage{tikz}
\usepackage{listings}
\usepackage{microtype}
\usepackage[hidelinks]{hyperref}
\hypersetup{pdftitle={What extraction cannot slim, and whether function injection can: a pre-registered study of 580 Java repositories},
  pdfauthor={Daniel Sagenschneider}, pdfsubject={Software engineering: empirical study of long-method refactoring},
  pdfkeywords={refactoring, Extract Method, Long Method, function injection, dependency injection, pre-registered study, mining software repositories, Java}}
\usepackage{caption}
\newcommand{\nDrawn}{757}
\newcommand{\nUsable}{580}
\newcommand{\nExcluded}{177}
\newcommand{\nError}{0}
\newcommand{\nScanned}{580}

\newcommand{\sampleFracHigh}{46\%}
\newcommand{\nMethods}{3{,}026{,}902}
\newcommand{\nLong}{168{,}466}
\newcommand{\longShare}{5.6\%}
\newcommand{\nUnparsed}{173}
\newcommand{\nSlimmable}{105{,}740}
\newcommand{\nLocked}{24{,}171}
\newcommand{\nRecord}{3{,}619}
\newcommand{\nNotSlimmable}{34{,}936}
\newcommand{\nSlims}{129{,}911}
\newcommand{\slimsShare}{77.1\%}
\newcommand{\nStuck}{38{,}555}
\newcommand{\stuckShare}{22.9\%}

\newcommand{\slimCostMedian}{3}
\newcommand{\slimCostQone}{2}
\newcommand{\slimCostQthree}{6}

\newcommand{\nStuckPlanned}{21{,}350}

\newcommand{\popStuckPlanned}{55.4\%}

\newcommand{\nStuckPlannedC}{24{,}346}

\newcommand{\ciStuckPlannedC}{63.2\% (95\% CI 60.0 to 66.4)}

\newcommand{\pooledCatThreeC}{22.1\%}
\newcommand{\popCatThreeC}{22.2\%}

\newcommand{\hiCatThreeC}{24.4}

\newcommand{\medianCatThreeC}{17.6\%}
\newcommand{\reposCatThreeC}{470}
\newcommand{\ciCatThreeC}{22.2\% (95\% CI 20.3 to 24.4)}
\newcommand{\tierHighCatThreeC}{22.5\%}
\newcommand{\tierMidCatThreeC}{21.5\%}
\newcommand{\tierLowCatThreeC}{22.4\%}
\newcommand{\nImproved}{35{,}656}

\newcommand{\pooledImproved}{27.5\%}
\newcommand{\popImproved}{29.0\%}

\newcommand{\medianImproved}{21.5\%}

\newcommand{\ciImproved}{29.0\% (95\% CI 24.3 to 34.6)}

\newcommand{\nTied}{26{,}055}

\newcommand{\pooledTied}{20.1\%}
\newcommand{\popTied}{20.9\%}

\newcommand{\catOne}{17{,}132}
\newcommand{\catOneC}{17{,}132}
\newcommand{\catOneShare}{44.4\%}
\newcommand{\catOneCShare}{44.4\%}
\newcommand{\catTwo}{3{,}619}
\newcommand{\catTwoC}{3{,}619}
\newcommand{\catTwoShare}{9.4\%}
\newcommand{\catTwoCShare}{9.4\%}
\newcommand{\catThree}{7{,}533}
\newcommand{\catThreeC}{8{,}540}

\newcommand{\catThreeCShare}{22.2\%}

\newcommand{\catFourC}{9{,}264}

\newcommand{\catFourCShare}{24.0\%}

\newcommand{\nStuckNoPlanC}{14{,}209}

\newcommand{\loopShareStuckNoPlanC}{43.9\%}
\newcommand{\jumpsStuckNoPlan}{177}
\newcommand{\jumpsStuckNoPlanC}{483}

\newcommand{\nBoth}{111{,}564}
\newcommand{\splitSmallerSeam}{63.4\%}
\newcommand{\equalSeam}{32.1\%}
\newcommand{\splitLargerSeam}{4.5\%}

\newcommand{\splitLargerParts}{66.8\%}

\newcommand{\nWorse}{4{,}972}

\newcommand{\worseByOneShare}{84.1\%}

\newcommand{\notImprovedShare}{52.5\%}
\newcommand{\nNotImproved}{68{,}200}
\newcommand{\savingMedian}{2}
\newcommand{\savingQone}{1}
\newcommand{\savingQthree}{4}

\newcommand{\savingOneShare}{39.2\%}
\newcommand{\nImprovedStrict}{21{,}665}

\newcommand{\popImprovedStrict}{17.3\%}

\newcommand{\ciImprovedStrict}{17.3\% (95\% CI 14.6 to 20.1)}
\newcommand{\medianImprovedStrict}{12.0\%}

\newcommand{\topTenShare}{38.0\%}

\newcommand{\topTenCatThreeCMin}{16.9\%}
\newcommand{\topTenCatThreeCMax}{33.9\%}

\newcommand{\reposStuckNoCatThree}{124}

\newcommand{\reposStuckNoCatThreeShare}{26.4\%}

\newcommand{\testDroppedLong}{8{,}191}
\newcommand{\testDroppedLongShare}{4.9\%}

\newcommand{\testRepos}{80}
\newcommand{\testTopSegment}{test-applications}

\newcommand{\testSecondSegment}{internalClusterTest}

\newcommand{\testHOneHi}{24.8}

\newcommand{\testHOneCi}{22.7\% (95\% CI 21.0 to 24.8)}
\newcommand{\testHTwoCi}{29.5\% (95\% CI 24.9 to 35.1)}

\newcommand{\planInCatOneShare}{76.0\%}

\newcommand{\planInCatTwoShare}{77.0\%}

\newcommand{\nNotConventional}{17{,}804}
\newcommand{\rescueOfNotConventional}{48.0\%}

\newcommand{\rescueOfNotConventionalMedian}{50.0\%}
\newcommand{\rescueOfNotConventionalCi}{46.9\% (95\% CI 43.2 to 50.9)}
\newcommand{\nCatThreeLost}{189}

\newcommand{\catThreeCRestoredShare}{22.6\%}
\newcommand{\hOneBarGap}{0.6}
\newcommand{\shapeNImproved}{35{,}656}

\newcommand{\shapeImprovedSequenceReturns}{17.1\%}

\newcommand{\shapeImprovedAnySequence}{96.4\%}

\newcommand{\shapeTiedAnySequence}{97.7\%}

\newcommand{\shapeNRescued}{8{,}540}

\newcommand{\shapeRescuedSequenceReturns}{10.0\%}

\newcommand{\shapeRescuedAnyBranching}{64.2\%}
\newcommand{\shapeRescuedReturnsOfBranching}{29.9\%}
\newcommand{\nSeqWay}{37{,}430}
\newcommand{\popSeqWay}{23.8\%}
\newcommand{\ciSeqWay}{23.8\% (95\% CI 19.2 to 29.8)}

\newcommand{\popSeqWayStrict}{14.8\%}
\newcommand{\ciSeqWayStrict}{14.8\% (95\% CI 12.2 to 18.0)}

\newcommand{\nExceptions}{16}

\newcommand{\exceptionsShare}{0.012\%}

\newcommand{\scanDate}{15 November 2026}
\newcommand{\longestMethodLines}{5{,}976}
\newcommand{\longestMethodRepo}{plantuml/\allowbreak plantuml}

\newcommand{\planSeed}{20261101}

\newcommand{\planAsOf}{1 December 2020}
\newcommand{\planMinFunctions}{50}
\newcommand{\planMinFiles}{15}
\newcommand{\samplerVersion}{0.3.0}
\newcommand{\slimVersion}{0.5.4}

\newcommand{\planHOneShare}{25\%}
\newcommand{\planHOneMinimum}{30}
\newcommand{\planResamples}{10{,}000}
\newcommand{\planLevel}{95\%}
\newcommand{\priorStuck}{119}
\newcommand{\priorCatThree}{12}
\newcommand{\priorCatThreeShare}{10.1\%}
\newcommand{\priorSlimmed}{327}
\newcommand{\priorImproved}{95}
\newcommand{\priorImprovedShare}{29\%}

\newcommand{\versPrevious}{0.5.2}

\newcommand{\versDiffering}{285}
\newcommand{\versDifferingMethods}{4{,}307}

\newcommand{\exaLines}{32}
\newcommand{\exaInjTotal}{4}
\newcommand{\exaInjFunctions}{2}
\newcommand{\exaInjEdges}{2}

\newcommand{\exaInjFunOneDeclares}{0}
\newcommand{\exaInjFunTwoDeclares}{0}

\newcommand{\exaSinewObstacle}{jumps}

\newcommand{\exbLines}{40}
\newcommand{\exbInjTotal}{7}
\newcommand{\exbInjFunctions}{3}
\newcommand{\exbInjEdges}{2}

\newcommand{\exbCombSum}{5}
\newcommand{\exbCombFunctions}{2}
\newcommand{\exbCombEdges}{1}

\newcommand{\exbCombDeclares}{\texttt{records}, \texttt{skipped}}
\newcommand{\exbInjFunOneDeclares}{0}
\newcommand{\exbInjFunTwoDeclares}{1}
\newcommand{\exbInjFunThreeDeclares}{1}
\newcommand{\exbCombFunOneDeclares}{0}
\newcommand{\exbCombFunTwoDeclares}{2}
\newcommand{\exbSinewCost}{9}
\newcommand{\exbSinewWidth}{4}
\newcommand{\exbSinewPieces}{2}

\newcommand{\exbSinewCostPlusOne}{10}
\newcommand{\exbSavingWrapped}{5}
\newcommand{\exbSavingPlain}{4}

\newcommand{\exbPieceOneInputs}{\texttt{rows}, \texttt{options}, \texttt{records}}
\newcommand{\exbPieceOneOutputs}{\texttt{skipped}}
\newcommand{\exbPieceOneInputsN}{3}
\newcommand{\exbPieceOneOutputsN}{1}

\newcommand{\exbPieceTwoInputs}{\texttt{options}, \texttt{records}, \texttt{summary}}
\newcommand{\exbPieceTwoOutputs}{nothing}
\newcommand{\exbPieceTwoInputsN}{3}
\newcommand{\exbPieceTwoOutputsN}{0}

\newcommand{\exTables}{3{,}030}
\newcommand{\exTablesShare}{32.7\%}
\newcommand{\exIndivisibleShare}{49.4\%}
\newcommand{\exDataShare}{36.9\%}
\newcommand{\exControlShare}{13.6\%}
\newcommand{\exControlContainedShare}{91\%}
\newcommand{\exRegrouped}{220}
\newcommand{\exLimitSix}{4.3\%}
\newcommand{\exLimitTen}{12.2\%}
\newcommand{\exRepos}{331}
\newcommand{\exHotRepos}{32}
\newcommand{\exHotShare}{38\%}
\newcommand{\exHotLongShare}{20\%}
\newcommand{\exCogCatFour}{46.1}
\newcommand{\exCogRescued}{45.5}
\newcommand{\exCogConventional}{59.3}
\newcommand{\exCogSlimmable}{35.0}
\newcommand{\exCogPerLineCatFour}{0.36}
\newcommand{\exCogPerLineRescued}{0.33}
\newcommand{\exCogPerLineConventional}{0.47}
\newcommand{\exReread}{9{,}262}
\newcommand{\exIndivisible}{4{,}580}
\newcommand{\exIndivisibleRepos}{280}
\newcommand{\exData}{3{,}420}

\newcommand{\exControl}{1{,}264}

\newcommand{\exSwitches}{1{,}528}
\newcommand{\exSwitchesShare}{16.5\%}
\newcommand{\exSwitchOptions}{1{,}374}

\newcommand{\exSwitchShort}{1{,}028}
\newcommand{\exSwitchFlat}{346}
\newcommand{\exSwitchCaseOver}{154}
\newcommand{\exIfs}{750}
\newcommand{\exIfsShare}{8.1\%}
\newcommand{\exIfDispatch}{375}
\newcommand{\exIfPasses}{267}
\newcommand{\exIfLeaf}{75}
\newcommand{\exIfLeafShare}{0.8\%}
\newcommand{\exIfNonDispatch}{375}
\newcommand{\exIfPassOrJump}{300}
\newcommand{\exIfPassOrJumpShare}{80\%}
\newcommand{\exIfDenseLengthLo}{141}
\newcommand{\exIfDenseLengthHi}{186}
\newcommand{\exIfDenseCplLo}{0.53}
\newcommand{\exIfDenseCplHi}{0.59}
\newcommand{\exControlLoop}{852}
\newcommand{\exControlTry}{200}
\newcommand{\exControlSwitch}{97}
\newcommand{\exControlTop}{81}
\newcommand{\exControlContained}{1{,}149}
\newcommand{\exControlTopReturn}{55}
\newcommand{\exCogPerLineLoopLo}{0.57}
\newcommand{\exCogPerLineLoopHi}{0.63}
\newcommand{\exCogPerLineFlat}{0.12}
\newcommand{\exCogPerLineShort}{0.21}
\newcommand{\exLimitSixN}{398}

\newcommand{\exLimitTenN}{1{,}132}
\newcommand{\exLimitSixConventionalN}{328}

\newcommand{\exDataPlan}{1{,}132}
\newcommand{\exDataPlanShare}{33\%}
\newcommand{\exDataBody}{1{,}134}
\newcommand{\exDataBodyShare}{33\%}
\newcommand{\exDataNone}{1{,}154}
\newcommand{\exDataNoneShare}{34\%}
\newcommand{\exDataNoneCogPerLine}{0.46}
\newcommand{\exDataNoneLength}{91}
\newcommand{\exDispersion}{7.8}
\newcommand{\exTopTenShare}{38.8\%}
\newcommand{\exTopTenSizeShare}{38.3\%}

\newcommand{\exArrayTopTen}{87\%}
\newcommand{\exArrayTopTenSize}{18\%}
\newcommand{\exArrayDispersion}{11.5}
\newcommand{\exFlatDispersion}{1.8}
\newcommand{\exIfPassDispersion}{1.0}

\newcommand{\exNestedTryTopTen}{88\%}
\newcommand{\exRateOverall}{5.5\% (95\% CI 4.3 to 7.1)}

\newcommand{\exLitCatchmentArray}{80\%}
\newcommand{\exLitCatchmentConcat}{62\%}
\newcommand{\exLogicJunctionCase}{90\%}
\newcommand{\exLogicJunctionTry}{83\%}
\newcommand{\exBudgetRepos}{\texttt{OpenLiberty/\allowbreak open-liberty}, \texttt{apache/\allowbreak hive}}

\newcommand{\fitwidth}[1]{\resizebox{\ifdim\width>\textwidth\textwidth\else\width\fi}{!}{#1}}
\newcommand{\sig}[1]{\texttt{#1}}
\newcommand{\Hone}{H1}
\newcommand{\Htwo}{H2}

\title{\bfseries What extraction cannot slim, and whether function injection can\\
A pre-registered study of \nUsable{} Java repositories}

\newcommand{\thedoi}{10.5281/zenodo.23991506}
\newcommand{\doiline}{\ifdefined\thedoi\quad\small\texttt{doi:\thedoi}\fi}

\newcommand{\regCommit}{bcb5d48}

\author{Daniel Sagenschneider\\\small OfficeFloor\\\small\texttt{daniel@officefloor.net}\doiline}

\date{November 2026}

\begin{document}
\maketitle

\begin{abstract}
\noindent
A method of 26 code lines or more reads as a Long Method, and Extract Method
is the usual fix. Some long methods have no single-exit fix: a return cannot
leave an extracted piece, a piece hands back one value, and some methods are
knotted so that every cut is blocked. Function injection, as the OfficeFloor
framework does it, cuts differently: each piece becomes a function the
framework injects and sequences, an early return ends a function, and a
function hands on several values. The framework's author pre-registered a
test of whether it takes apart what extraction cannot. A stratified random
sample of \nUsable{} open-source Java repositories, each pinned before AI
coding assistants were in general use, was scanned by released tools whose
cost models were fixed before the sample was drawn: \nMethods{} methods,
\nLong{} of them long. Single-exit extraction cannot slim \stuckShare{} of the long methods.
The registered hypothesis, that at least a quarter of those stuck methods are
ones only function injection takes apart, is refuted: \ciCatThreeC{}. Nearly
half of the stuck methods are early-return methods the plan counts as
conventionally fixable, and the split takes three in four of those apart as
they stand. Where extraction would do, the split chooses injection and costs
less for \popImproved{} against extraction wrapped as one function and
\popImprovedStrict{} against plain extraction, almost always as a straight
sequence of functions. An exploratory reading of the quarter of the stuck
methods that neither takes apart finds a third of it tables and literals,
long for their data rather than their logic, and most of the rest blocked by
a jump inside a loop, by a nested statement too large to cut, or by values
too many for any cut. Three registered studies are proposed: the classes as
rules in the tools, a refinement per class, and the same measures on code
written with AI assistants.
\end{abstract}

\section{Introduction}

Fowler catalogued the Long Method and named its fix, Extract Method
\cite{fowler1999, fowler2018}. An earlier study of this author's found that
the length at which reviewers call a method a Long Method is 26 code lines,
lines that hold code with blank and comment lines not counted
\cite{longmethod}. A tool, SinewSlim, followed: for every method of 26 lines
or more it searches for a set of extractions that brings the method, and
every extracted piece, to 25 lines or fewer, pricing each cut by what crosses
it and taking the segmentation whose widest cut is narrowest. Tools that find
Extract Method opportunities are not new \cite{tsantalis2011, haas2016};
SinewSlim's difference is that it prices every cut and says when no cut is
feasible, and that is where this paper starts. Some long methods cannot be
slimmed by extraction into single-exit pieces at all. A \sig{return} cannot
leave such a piece, so a method of guard clauses blocks every cut that
contains one. An extracted piece hands back one value, so a run of statements
that produces two needs a record to carry them, at a cost of its own. Some
methods are knotted so tightly that every cut needs more parameters than a
reader should follow. These methods are called \emph{stuck} here, and stuck
means stuck under extraction as SinewSlim models it: one exit, one output and
at most five parameters a piece. The first finding of this paper is the count
of them: on \nUsable{} repositories, \stuckShare{} of the long methods are
stuck, between a fifth and a quarter.

Function injection cuts a method differently. In OfficeFloor, a framework the
author created, a function is an ordinary Java method whose arguments the
framework supplies, as dependency injection does \cite{fowler2004}, and whose
next function the framework chooses: the developer writes the pieces and the
wiring between them, and the framework sequences them. An early return
becomes a flow from its function to the method's end; an \sig{if} can become a
decision with a flow per branch; a function hands on several values, because
the next function asks the framework for them; the instrument used here lets
a function take up to five such values as parameters. So a method extraction
cannot slim might still come apart as functions. The author expected it would, often. That expectation is exactly
what a reader should not take on trust from the framework's author, and this
paper is the pre-registered test of it: the sample's seed, the hypotheses,
the bar and the rule for reading the result were committed before any
repository of the sample was read, every cost comes from released products
whose cost models were fixed at or before registration, and every figure is reported
whichever way it points. Each repository of the sample is pinned to its last
commit before \planAsOf{}, six months before GitHub Copilot's technical
preview of 29 June 2021 \cite{copilot2021}, so that no method in the sample
was written with the AI coding assistants in general use since.

\paragraph{What the paper reports.} Four things. The count of the long
methods single-exit extraction cannot slim, by cause, on a seeded random
sample of \nUsable{} repositories. The registered test of whether function
injection takes apart a quarter of them, which it fails: \ciCatThreeC{}, the
bar \hOneBarGap{} points above the interval, with a reading, not registered,
of where the shortfall sits. An estimate of how often, where extraction would
do, the split a developer would make chooses injection and costs less under
cost models fixed at or before registration, with the shape of the plans it
chooses and the mechanism, derived from the two cost models, that decides
them. And an exploratory reading of the quarter of the stuck methods that no
split takes apart, from which three further studies are proposed. Every
reason the split fails and every method where it does worse are in
the replication package, from which every figure here is regenerated by
script.

\section{Background}\label{sec:background}

\paragraph{Extraction, and its cost.} SinewSlim reads a method's facts (its
statements, the values each declares, reads and writes, the jumps that leave
each run of statements) and searches for the segmentation $S$ into contiguous
runs whose extraction leaves every piece and the remainder at or under the
target. A piece $p \in S$ is priced by its \emph{seam}, what crosses its
boundary: its $i_p$ inputs, passed as parameters, and its $o_p$ outputs,
handed back; and the piece itself costs 1, for its name and its call. The
extraction cost of a segmentation is
\begin{equation}\label{eq:extraction}
  C_{\mathrm{ext}}(S) \;=\; \sum_{p \in S}
    \bigl(\, \underbrace{i_p + r(o_p)}_{\text{seam of } p} \,+\, 1 \,\bigr),
  \qquad
  r(o) = \begin{cases} o, & o \le 1,\\ 3o, & o \ge 2 \text{ (a record)}. \end{cases}
\end{equation}
A run is feasible with $i_p \le 5$, $o_p \le 1$ and no jump leaving it other
than a trailing return. The inputs count every value a piece reads that was
set before it, the method's own parameters included, except a value declared
before the piece with no value or a literal one and first used inside it,
which moves into the piece, unless the piece is inside a loop that writes
the value, where the piece reads what the previous iteration left and the
value is an input; the remainder, what is left in the method, is not
priced. The search takes the segmentation with the
smallest worst seam $\max_{p} (i_p + r(o_p))$, then the fewest pieces $|S|$,
then the smallest $C_{\mathrm{ext}}$: it minimises the widest cut first and
the total last, so SinewSlim's segmentation is not in general the one of
least cost, a point Section~\ref{sec:threats} returns to. A method is
\emph{slimmable} when a segmentation exists; \emph{slimmable but locked} when
every piece and the remainder are bound to the class's fields or methods, so
that a split moves nothing out of the class; \emph{slimmable only with a
record} when a piece must hand back several values ($o_p \ge 2$, priced at
$3o_p$); and \emph{not slimmable} otherwise, with the obstacle named. The
last two statuses are what this paper calls stuck.

\paragraph{The injection split.} OfficeFloorSlim, the second product, reads
the same facts and searches for a partition $P$ of the method's control flow
into functions $F$ joined by edges $E$, each function at or under the target,
with every line in some function. Its cost model was decided before any of
its figures was set beside SinewSlim's:
\begin{equation}\label{eq:injection}
  C_{\mathrm{inj}}(P) \;=\; \sum_{f \in F} \bigl(\, d_f + 1 \,\bigr) \;+\; |E|,
\end{equation}
where $d_f$, the function's seam, counts the values computed inside the
method that $f$ declares as its parameters, so that a value used by several
later functions costs 1 in each function that declares it; values from
outside the method, its parameters and whatever the framework injects, are
free and not counted, and what a function returns to the framework is not
counted either; a value declared earlier with no value or a literal one and
first used inside the function moves into it, as into SinewSlim's pieces,
and is not declared, unless, as there, the function is inside a loop that
writes it. Each function costs 1 and each edge, a flow between
functions, costs 1; a function may carry at most five values, $d_f \le 5$. A
\sig{return} before a function's end is one flow to the method's end,
however many such returns the function holds; a \sig{throw} becomes an
escalation; and an \sig{if} or \sig{switch} may become a decision function
in $F$ whose branches are edges in $E$. Loops, lambda bodies and labelled statements are never cut
between functions. The injection split chooses among feasible partitions by
SinewSlim's own rule: the smallest worst seam $\max_f d_f$, then the fewest
functions, then the smallest $C_{\mathrm{inj}}$. Its worst seam is the
counterpart of SinewSlim's widest piece, with one difference that matters for
reading them side by side: SinewSlim's $i_p$ counts the method's own
parameters when a piece reads them, and $d_f$ does not, since the framework
injects them.

\paragraph{The combined split.} No OfficeFloor developer would wire a
ten-line calculation through the framework as functions and flows when
extracting a method would do. The combined split is what a developer would
do: functions where the flow divides, extracted methods inside them, its
rule settled on the day of registration, before the sample was drawn. Its
outer search is the injection split's; a function whose body is over the
target is allowed when SinewSlim's search, under its defaults, slims that
body as a method of its own, its declared values its parameters. A body
SinewSlim slims only with a record is refused. Writing $S_f$ for SinewSlim's
segmentation of the body of $f$ (empty when the body is at or under the
target), a plan is priced by the \emph{sum}
\begin{equation}\label{eq:combined}
  C_{\mathrm{comb}}(P) \;=\; C_{\mathrm{inj}}(P) \;+\; \sum_{f \in F} C_{\mathrm{ext}}(S_f),
\end{equation}
the injection total of~\eqref{eq:injection} plus the extraction cost
of~\eqref{eq:extraction} of the pieces inside the functions. Unlike the two
splits it is built from, the combined split chooses by the total: the plan
with the smallest $C_{\mathrm{comb}}$; on a tie, the fewer pieces
$\sum_f |S_f|$, then the fewer functions $|F|$. The whole method as one
function, slimmed by SinewSlim, is always a candidate: with $|F| = 1$, no
edge and nothing carried, $C_{\mathrm{inj}} = 1$ and
$C_{\mathrm{comb}} = C_{\mathrm{ext}}(S) + 1$. So wherever SinewSlim slims a
method the combined split has a plan, and
$\min_P C_{\mathrm{comb}}(P) \le C_{\mathrm{ext}}(S) + 1$. That holds by
construction and is not a finding; a one-function plan is reported as
\emph{extraction only} and counted as SinewSlim's.

\paragraph{What the two models price.} Equations~\eqref{eq:extraction}
and~\eqref{eq:injection} price the same thing, data crossing a boundary, by
different rules, and the rules say in advance where each is cheaper.

Extraction prices a computed value twice: once on its way out of the piece
that made it, as an output, and once on its way into each piece that reads
it, as an input. A piece that hands back two or more values prices each
three times, as a field of a record. The remainder is free: a value that
flows back into what stays in the method costs nothing on arrival. A
method's own parameters are priced like any other input, once into each
piece that reads them. And each piece costs 1, for its name and its call.

Injection prices a computed value once, in each function that declares it
as a parameter, and nothing on its way out. A method's parameters are never
priced, since the framework injects them wherever they are read, and what a
function returns to the framework is not priced either. Each function costs
1, and each flow between functions costs 1, whether or not anything crosses
it.

So a boundary that one computed value crosses costs the same both ways, an
output and an input against a declared value and a flow, and a chain of $k$
single-value stages costs within 1 of the same under both. The models part
in four places. A flow is cheaper where several values cross one boundary,
each priced once against twice or three times; where a method's parameters
are read past the first cut, which extraction prices and injection does not;
and where what the last piece hands back is the method's own result, which
extraction prices as an output and injection returns for free. It is dearer
where a boundary carries nothing, since the flow still costs 1 and an empty
seam costs 0. The second of these is what the example \sig{build} below
shows. Which of them the long methods of the sample contain is what
Section~\ref{sec:results} measures.

\paragraph{Two worked examples.} Figure~\ref{fig:examples} gives two methods
written for this paper and priced by the pinned instrument; the comments in
the code mark the cuts it reports and name each piece and function as the
equations below do: \sig{EXTRACT} for a piece SinewSlim would extract,
\sig{FUNCTION} for a function of the combined split, and \sig{INJECTION}
for a function of the injection split alone where that differs. The costs
below are the instrument's, set out term by term under
Equations~\eqref{eq:extraction} to~\eqref{eq:combined}.

The first, \sig{describe}, is \exaLines{} code lines of guard clauses, and
the point of it is that one model has a price and the other has none. Every
run of more than a line or two contains a \sig{return}, so no single-exit
piece is feasible and Equation~\eqref{eq:extraction} has nothing to sum.
The injection split makes two functions, the guards and the rest, each
declaring nothing; the guards' early returns are one flow to the method's
end, the returns of \sig{price} are its own end and cost nothing, and one
flow joins the two:
\begin{align*}
  C_{\mathrm{ext}}(\sig{describe}) &\;\text{ has no value: no feasible segmentation } S
    \text{ (SinewSlim's obstacle: \emph{\exaSinewObstacle}),}\\
  C_{\mathrm{inj}}(\sig{describe}) &\;=\;
    \underbrace{(\exaInjFunOneDeclares{} + 1)}_{\sig{check}} \;+\;
    \underbrace{(\exaInjFunTwoDeclares{} + 1)}_{\sig{price}} \;+\;
    \underbrace{\exaInjEdges{}}_{|E|} \;=\; \exaInjTotal{}.
\end{align*}
Its jumps ignored, the method would slim, so the plan puts it in category 1
of the four defined after the examples, conventionally fixable, and credits the split with it only in the
post hoc reading.

The second, \sig{build}, is \exbLines{} lines in three stages: the first
produces \sig{records} and \sig{skipped}, read by what follows, and the
parameter \sig{options} is read throughout. SinewSlim extracts the two loops,
\sig{parseRows} taking \exbPieceOneInputs{} and handing back
\exbPieceOneOutputs{}, and \sig{summarise} taking \exbPieceTwoInputs{} and
handing back \exbPieceTwoOutputs{}:
\begin{equation*}
  C_{\mathrm{ext}}(\sig{build}) \;=\;
    \underbrace{(\exbPieceOneInputsN{} + r(\exbPieceOneOutputsN{}) + 1)}_{\sig{parseRows}} \;+\;
    \underbrace{(\exbPieceTwoInputsN{} + r(\exbPieceTwoOutputsN{}) + 1)}_{\sig{summarise}} \;=\; \exbSinewCost{},
  \qquad
  C_{\mathrm{comb}}(\text{one function}) \;=\; \exbSinewCost{} + 1 \;=\; \exbSinewCostPlusOne{}.
\end{equation*}
The widest piece is \sig{parseRows}, seam \exbSinewWidth{}. The injection
split alone, choosing by worst seam first, cuts at both stage boundaries so
that no function declares more than one value, and pays a flow at each. Its
three functions are marked \sig{INJECTION} in the figure: \sig{prepare}
holds the two declarations and declares nothing; \sig{tally} holds both
loops and declares \sig{records} alone, since \sig{skipped}, declared with a
literal and first used inside it, moves in, as it moves into SinewSlim's
\sig{parseRows}, which hands it back; and \sig{report} builds the report
and declares \sig{summary}:
\begin{equation*}
  C_{\mathrm{inj}}(\sig{build}) \;=\;
    \underbrace{(\exbInjFunOneDeclares{} + 1)}_{\sig{prepare}} \;+\;
    \underbrace{(\exbInjFunTwoDeclares{} + 1)}_{\sig{tally}} \;+\;
    \underbrace{(\exbInjFunThreeDeclares{} + 1)}_{\sig{report}}
    \;+\; \underbrace{\exbInjEdges{}}_{|E|} \;=\; \exbInjTotal{}.
\end{equation*}
The combined split, choosing by the total, cuts once, before the second
stage; the second function declares \exbCombDeclares{}, and both bodies are
at or under the target, so no piece is extracted inside either:
\begin{equation*}
  C_{\mathrm{comb}}(\sig{build}) \;=\;
    \underbrace{(\exbCombFunOneDeclares{} + 1)}_{\sig{parse}} \;+\;
    \underbrace{(\exbCombFunTwoDeclares{} + 1)}_{\sig{summarise}} \;+\;
    \underbrace{\exbCombEdges{}}_{|E|} \;+\;
    \underbrace{0}_{\sum_f C_{\mathrm{ext}}(S_f)} \;=\; \exbCombSum{},
\end{equation*}
a saving of \exbSavingWrapped{} against extraction wrapped as one function
and \exbSavingPlain{} against plain extraction. The \exbSavingPlain{} is
the three parameters, \sig{options} priced into each piece and \sig{rows}
into the first, which the split never prices; \sig{summary}, which
extraction separates from the loop that fills it and the split keeps with
it; and \sig{records}, an input to both pieces but declared by one
function; against those five the split pays 1 for its flow.

\begin{figure}[tbp]
\begin{minipage}[t]{0.47\textwidth}
\begin{lstlisting}[firstline=9,lastline=44]
    public String describe(Order order, Customer customer, Catalog catalog) {
        // FUNCTION check: a return ends the flow, else next
        if (order == null) {
            return "no order";
        }
        if (order.getLines().isEmpty()) {
            return "empty order";
        }
        if (customer == null) {
            return "no customer";
        }
        if (!customer.isActive()) {
            return "inactive customer";
        }
        // END FUNCTION: passes nothing on
        // FUNCTION price: declares nothing
        int total = catalog.totalOf(order.getLines());
        if (total == 0) {
            return "free";
        }
        int discount = customer.discountFor(total);
        if (discount > total) {
            return "invalid discount";
        }
        int due = total - discount;
        String currency = customer.getCurrency();
        if (currency == null) {
            return "no currency";
        }
        String note = customer.noteFor(due);
        if (note == null) {
            return currency + " " + due;
        }
        return currency + " " + due + " (" + note + ")";
        // END FUNCTION: returns the String
    }
\end{lstlisting}
\end{minipage}\hfill
\begin{minipage}[t]{0.5\textwidth}
\begin{lstlisting}[firstline=48,lastline=101]
    public Report build(List<String> rows, Options options) {
        // FUNCTION parse: declares nothing
        // INJECTION prepare: declares nothing
        List<Record> records = new java.util.ArrayList<>();
        int skipped = 0;
        // END INJECTION: passes on records
        // INJECTION tally: declares records
        // EXTRACT parseRows(rows, options, records)
        for (String row : rows) {
            String[] cells = row.split(options.separator());
            if (cells.length < 3) {
                skipped++;
                continue;
            }
            Record record = new Record();
            record.setKey(cells[0].trim());
            record.setValue(Integer.parseInt(cells[1].trim()));
            record.setUnit(cells[2].trim());
            record.setSource(options.source());
            records.add(record);
        }
        // END EXTRACT: returns skipped
        // END FUNCTION: passes on records, skipped
        // FUNCTION summarise: declares records, skipped
        Summary summary = new Summary();
        // EXTRACT summarise(options, records, summary)
        for (Record record : records) {
            if (record.getValue() < options.minimum()) {
                summary.addInvalid(record.getKey());
                continue;
            }
            summary.addTotal(record.getValue());
            summary.addUnit(record.getUnit());
            summary.addSource(record.getSource());
            summary.incrementCount();
        }
        // END EXTRACT: returns nothing
        summary.setSkipped(skipped);
        // END INJECTION: passes on summary
        // INJECTION report: declares summary
        Report report = new Report();
        report.setTitle(options.title() + ": " + summary.getCount() + " records");
        report.setTotal(summary.getTotal());
        report.setUnits(summary.getUnits());
        report.setSources(summary.getSources());
        report.setInvalid(summary.getInvalid());
        report.setSkipped(summary.getSkipped());
        report.setGenerated(System.currentTimeMillis());
        report.setFormat(options.format());
        report.finish();
        return report;
        // END INJECTION: returns report
        // END FUNCTION: returns report
    }
\end{lstlisting}
\end{minipage}
\caption{Two worked examples, priced by the pinned instrument; the comments mark the cuts it reports, named as in the equations of the text, and are not code lines. Left, \sig{describe}: stuck under single-exit extraction (obstacle \emph{\exaSinewObstacle}); the injection split and the combined split both make the \exaInjFunctions{} functions marked \sig{FUNCTION}, with \exaInjEdges{} flows, carrying nothing, total \exaInjTotal{}. Right, \sig{build}: SinewSlim extracts the \exbSinewPieces{} pieces marked \sig{EXTRACT}, each to the method named, at a cost of \exbSinewCost{}; the injection split alone makes the \exbInjFunctions{} functions marked \sig{INJECTION} with \exbInjEdges{} flows, total \exbInjTotal{}; the combined split makes the \exbCombFunctions{} functions marked \sig{FUNCTION} with \exbCombEdges{} flow, the second declaring \exbCombDeclares{}, total \exbCombSum{}.}
\label{fig:examples}
\end{figure}

\paragraph{The categories.} A stuck method is put in one category, tested
conventional first, so that function injection is credited only with what
nothing simpler fixes: category 2 if a record slims it; else category 1 if
SinewSlim's defaults slim it once its jumps are ignored, the guard-clause
methods a developer could restructure; else category 3 if the OfficeFloor
column has a plan; else category 4, which needs a redesign. Category 3 is
the rescue, and the plan's hypothesis reads it under the combined split.
Table~\ref{tab:glossary} collects the terms used from here on.

\begin{table}[t]
\centering\small
\caption{The terms.}
\label{tab:glossary}
\begin{tabular}{>{\raggedright\arraybackslash}p{3.3cm}>{\raggedright\arraybackslash}p{11.9cm}}
\toprule
Term & Meaning \\
\midrule
Code line & A line holding code; blank and comment lines are not counted \\
Long method & A method of 26 code lines or more; the target for any part is 25 \\
Piece, remainder & An extracted run of statements, and what stays in the method \\
Seam, width & What crosses a piece's boundary, inputs plus outputs, Eq.~\eqref{eq:extraction}; the width of a segmentation is its worst seam \\
Record & A piece handing back two or more values, priced at 3 each \\
Slimmable, locked & SinewSlim has a segmentation; locked when it moves nothing out of the class \\
Stuck & Slimmable only with a record, or not slimmable: no single-exit segmentation \\
Function, flow & A piece of the injection split, and an edge between two of them, Eq.~\eqref{eq:injection} \\
Declares, worst seam & The computed values a function takes as parameters; the most any function declares \\
Decision, escalation & A function made of an \sig{if} or \sig{switch}, whose branches are flows; a \sig{throw} \\
Injection split & Functions and flows alone, chosen by worst seam first \\
Combined split & Functions where the flow divides, SinewSlim's pieces inside them, chosen by the sum, Eq.~\eqref{eq:combined} \\
Extraction only & A combined plan of one function: SinewSlim's segmentation, plus 1 \\
Improved, tied & Where SinewSlim slims a method: the combined split chooses injection and its sum is below, or equal to, SinewSlim's cost plus 1 \\
Rescued & A stuck method in category 3: a plan under the OfficeFloor column and no conventional fix \\
Categories 1 to 4 & Of a stuck method, tested in the order 2, 1, 3, 4: a record slims it; extraction slims it once its jumps are ignored; only the split has a plan; nothing has a plan \\
Sequence, branching & A plan with no decision function, or with one; either may have early returns \\
\bottomrule
\end{tabular}
\end{table}

\paragraph{Related work.} Finding Extract Method opportunities is a settled
line of work. JDeodorant finds them by slicing \cite{tsantalis2011}, JExtract
by the dependencies among a method's statements \cite{silva2014}, SEMI by the
functional relevance of neighbouring statements \cite{charalampidou2017},
GEMS by a learned model \cite{xu2017}, and Haas and Hummel by a scored search
over long methods \cite{haas2016}. Maruyama's block-based slicing
\cite{maruyama2001} and Komondoor and Horwitz's semantics-preserving
extraction \cite{komondoor2000} take runs that are not contiguous or that
jump, which is where a single-exit piece fails; IDEs do the same for one
early return by handing back a flag the caller tests. JExtract is the
nearest in what it considers: it shares SinewSlim's candidate space,
contiguous runs with one exit, but scores a run by how far its variables,
types and packages differ from the rest of the method and returns the best
three, so it has no length target and no verdict when nothing is feasible.
Each of these proposes cuts; none counts what no cut can do across a corpus,
and that is the count this paper makes, under one tool's rules. Studies of refactoring as practised
find that most of it is done by hand and that Extract Method is among the
refactorings developers apply most \cite{murphyhill2012}; asked why,
developers name reuse and the readability of a long method
\cite{silva2016}; and RefactoringMiner finds the refactoring in commit
history at scale \cite{tsantalis2018}. The Long Method is among the smells
whose diffuseness across releases Palomba et al.\ measured
\cite{palomba2018}, and detectors for it, by threshold or learned, are
compared by Fontana et al.\ \cite{fontana2016}. A length threshold for a long
method is usually a statistic over a corpus \cite{lanza2006}; the 26 used
here is what reviewers call long \cite{longmethod}. Cognitive complexity
\cite{campbell2018}, read in Section~\ref{sec:explore}, measures how hard a
method is to follow rather than how long it is. Function injection has an
older relative in flow-based programming \cite{morrison2010}, where a program
is a graph of processes passing packets along connections, and a nearer one
in dependency injection \cite{fowler2004}; what OfficeFloor adds is that the
pieces are methods of ordinary classes and the sequencing is the framework's.
Pre-registration is newer to software engineering than to the sciences it
came from; registered reports, in which the plan is reviewed before the data
is read, are described by Ernst and Baldassarre \cite{ernst2023}, and this
study follows their form without a reviewer. The author knows of no prior
count of how many long methods a flow of functions can take apart that
extraction cannot.

\paragraph{What was known before the plan.} The combined split had been run
on one repository, OfficeFloor's own, written by the author, before the
hypotheses were written: it rescued \priorCatThree{} of \priorStuck{} stuck
methods (\priorCatThreeShare{}), and of the \priorSlimmed{} long methods
SinewSlim slims it chose injection and cost less for \priorImproved{}
(\priorImprovedShare{}). That one repository pointed against the rescue
hypothesis. The plan records the author's expectation all the same: that
code written by many hands, before AI assistants, would show more methods
that only injection takes apart than the framework's own code does. The
\priorImprovedShare{} is why the second question has no threshold: a share
set after seeing it would test nothing.

\section{Method}

The analysis plan was registered on 1 November 2026 with every decision
settled, before the sample was drawn, by commit \sig{\regCommit} of the
study's public repository. Every change since is a dated amendment that
records what had been seen when it was made. The plan, the pinned values,
the step scripts and the amendments are in that repository, named in the
ancillary data, and the plan as it stood at that commit is deposited with the
data (see Data availability). Everything below is as registered unless
marked post hoc or exploratory.

\paragraph{The sample.} An independent tool, open-source-repository-sampler
\samplerVersion{}, drew a seeded, stratified simple random sample of GitHub
repositories with the plan's seed, \planSeed{}, chosen as one nobody had
drawn before; the study's scripts refuse any other seed. The frame is Java
repositories created before \planAsOf{} in three popularity tiers, high (over
5{,}000 stars), mid (1{,}000 to 5{,}000) and low (200 to 999), drawn equally.
Each repository is pinned to the newest commit on its default branch's
first-parent history committed before \planAsOf{}, and judged there: at least
\planMinFunctions{} functions in at least \planMinFiles{} files, repositories
owned by the author or by OfficeFloor excluded. Table~\ref{tab:sample} gives
each tier's population, what was drawn and what was usable; population
estimates weight each repository by its stratum's population over its usable
count. The draw finished on 2 November 2026 and is committed: \nDrawn{} drawn,
\nUsable{} usable, \nExcluded{} excluded by the criteria, \nError{} lost to
errors.

\begin{table}[t]
\centering\small
\caption{The sample: each tier's population before the as-of date, what was drawn, what met the criteria, and the stratum weight (population over usable) used for population estimates.}
\label{tab:sample}
\begin{tabular}{llrrrrr}
\toprule
Tier & Stars & Population & Drawn & Usable & Excluded & Weight \\
\midrule
High & over 5{,}000 & 546 & 253 & 224 & 29 & 2.4 \\
Mid & 1{,}000 to 5{,}000 & 2{,}539 & 252 & 191 & 61 & 13.3 \\
Low & 200 to 999 & 8{,}818 & 252 & 165 & 87 & 53.4 \\
\midrule
All & & 11{,}903 & 757 & 580 & 177 & \\
\bottomrule
\end{tabular}

\end{table}

\paragraph{The instruments.} Every cost comes from released products; the
study computes none. OfficeFloorSlim \slimVersion{}, built on SinewSlim
0.2.4, was run once per repository at its pinned commit, from a jar whose
digest is pinned in the study. It reports, for every long method, SinewSlim's
assessment under SinewSlim's defaults (trigger 26, target 25), the injection
split and the combined split, each with its plan or its reason for having
none, and the method's category under each. The scope is production Java:
every \sig{.java} file, leaving out \sig{src/test}, \sig{src/it}, build
output, generated directories and any directory named \sig{test} or
\sig{tests}. The instrument was re-pinned three times before any figure of
the plan's measures had been computed, twice for speed and once for a fault
in an intermediate release, and a fourth time after the figures had been
computed and published, for a fault in SinewSlim up to 0.2.3: a value
carried round a loop was not an input of a piece inside the loop. The
release used differs from OfficeFloorSlim \versPrevious{}, whose figures the
first version of this paper printed, on \versDifferingMethods{} methods in
\versDiffering{} of the \nScanned{} repositories, and changes no hypothesis's
reading; the plan's amendments record each change and its evidence.

\paragraph{The measures and the analysis.} Per repository and pooled, the
plan fixes the long methods by SinewSlim's status and their slim cost, the
stuck methods and those each split has a plan for, the stuck methods by
category under each split, the worst seam, the number of parts and the
largest part side by side where both SinewSlim and the injection split have a
plan, each split's reasons where it has no plan, and, of the long methods
SinewSlim slims, those where the combined split chooses injection and its sum
is below SinewSlim's cost plus 1 (\emph{improved}) or equal to it
(\emph{tied}), with the saving. The two total costs are never compared: they
count different things, and the weight that would turn one into the other is
the very thing a sceptical reader would say was chosen to favour function
injection. The unit is the repository. Every interval is a percentile
bootstrap over repositories within strata, \planResamples{} resamples with
the plan's seed, at \planLevel{}; the bootstrap applies no finite population
correction, and the high tier's draw is \sampleFracHigh{} of its population,
so the intervals there are, if anything, wide. Each headline figure is
reported pooled, as a population estimate weighted by stratum with its
interval, per tier, and as the median of each repository's own share, so that
a few very large repositories cannot quietly carry a pooled figure.

\paragraph{The hypotheses.} Each is read from the \planLevel{} interval of
its population estimate. \Hone{}, the rescue, is confirmatory: category 3
under the combined split is at least \planHOneShare{} of the stuck methods
and at least \planHOneMinimum{} methods, a bar fixed before any scan;
supported if the whole interval is at or above \planHOneShare{} and the count
reaches the minimum, refuted if the whole interval is below, inconclusive
otherwise. \Htwo{}, the improvement, is an estimate with no threshold: the
share of the long methods SinewSlim slims where the combined split chooses
injection and costs less; ties are reported beside it and never counted,
because the tie-break favours injection.

\section{Results}\label{sec:results}

Every usable repository was scanned (\nScanned{} of \nUsable{}, none
failed) on \scanDate{}; \nUnparsed{} files could not be parsed and were left
out. Figure~\ref{fig:flow} gives the whole count at a glance and
Table~\ref{tab:status} the corpus. Of \nMethods{} methods with a body,
\nLong{} (\longShare{}) are long. SinewSlim slims \nSlims{} of them
(\slimsShare{}) with one return per piece, at a median slim cost of
\slimCostMedian{} (quartiles \slimCostQone{} and \slimCostQthree{}), the
cost being SinewSlim's total for the whole segmentation rather than any one
seam. The rest, \nStuck{} (\stuckShare{}), are stuck: \nRecord{} slim only
with a record and \nNotSlimmable{} not at all. The shares are alike across
the tiers.

\begin{figure}[t]
\centering
\fitwidth{%
\begin{tikzpicture}[
  every node/.style={font=\small},
  box/.style={draw, rounded corners=2pt, align=center, inner sep=3pt},
  >=stealth]
\node[box] (long) at (0,0) {\nLong{} long methods, \longShare{} of the \nMethods{} with a body};
\node[box] (slims) at (-4.4,-1.8) {SinewSlim slims \nSlims{} (\slimsShare)\\\footnotesize \nSlimmable{} slimmable, \nLocked{} locked to the class};
\node[box] (stuck) at (4.4,-1.8) {Stuck: \nStuck{} (\stuckShare)\\\footnotesize \nRecord{} only with a record, \nNotSlimmable{} not at all};
\node[box, text width=2.3cm] (imp) at (-7.0,-4.3) {Injection chosen and cheaper\\\nImproved{} (\pooledImproved)};
\node[box, text width=2.3cm] (tie) at (-4.4,-4.3) {Tied, settled by the tie-break\\\nTied{} (\pooledTied)};
\node[box, text width=2.3cm] (ext) at (-1.8,-4.3) {Extraction alone chosen\\\nNotImproved{} (\notImprovedShare)};
\node[box, text width=1.7cm] (c1) at (1.2,-4.3) {1: jumps ignored\\\catOneC{} (\catOneCShare)};
\node[box, text width=1.7cm] (c2) at (3.3,-4.3) {2: a record\\\catTwoC{} (\catTwoCShare)};
\node[box, text width=1.7cm] (c3) at (5.4,-4.3) {3: the rescue\\\catThreeC{} (\catThreeCShare)};
\node[box, text width=1.7cm] (c4) at (7.5,-4.3) {4: a redesign\\\catFourC{} (\catFourCShare)};
\draw[->] (long) -- (slims);
\draw[->] (long) -- (stuck);
\foreach \t in {imp,tie,ext} \draw[->] (slims) -- (\t);
\foreach \t in {c1,c2,c3,c4} \draw[->] (stuck) -- (\t);
\node[font=\footnotesize, align=center] at (-4.4,-5.7) {the combined split, where extraction would do};
\node[font=\footnotesize, align=center] at (4.4,-5.7) {the categories under the combined split, tested conventional first};
\end{tikzpicture}}
\caption{The count. Every long method of the \nUsable{} repositories, by SinewSlim's status; the methods it slims, by what the combined split chooses; the stuck methods, by category under the combined split. Shares are pooled over the sample.}
\label{fig:flow}
\end{figure}

\begin{table}[t]
\centering\small
\caption{The long methods by SinewSlim's status, per tier and pooled. The last two statuses are the stuck methods.}
\label{tab:status}
\begin{tabular}{lrrrrr}
\toprule
SinewSlim's status & High & Mid & Low & All & Share \\
\midrule
Methods with a body & 1{,}509{,}345 & 938{,}580 & 578{,}977 & 3{,}026{,}902 & \\
Long (26 code lines or more) & 70{,}918 & 58{,}536 & 39{,}012 & 168{,}466 & 5.6\% of methods \\
\midrule
Slimmable & 43{,}979 & 35{,}967 & 25{,}794 & 105{,}740 & 62.8\% \\
Slimmable, but locked to its class & 10{,}379 & 8{,}917 & 4{,}875 & 24{,}171 & 14.3\% \\
Slimmable only with a record (stuck) & 1{,}392 & 1{,}398 & 829 & 3{,}619 & 2.1\% \\
Not slimmable (stuck) & 15{,}168 & 12{,}254 & 7{,}514 & 34{,}936 & 20.7\% \\
\midrule
Stuck & 16{,}560 & 13{,}652 & 8{,}343 & 38{,}555 & 22.9\% of long \\
\bottomrule
\end{tabular}

\end{table}

\paragraph{The reach, and \Hone{}.} Table~\ref{tab:headline} gives every
headline figure and Table~\ref{tab:categories} the stuck methods by category
under each split. The combined split has a plan for \nStuckPlannedC{} of the
\nStuck{} stuck methods, \ciStuckPlannedC{} of the population, and the
injection split alone for \nStuckPlanned{} (\popStuckPlanned{}). But having a
plan is not the rescue. Tested conventional first, \catOne{} of the stuck
methods (\catOneShare{}) would slim by extraction once their jumps were
ignored, and \catTwo{} (\catTwoShare{}) with a record. What is left is
category 3, the methods only function injection takes apart: \catThreeC{}
under the combined split, \ciCatThreeC{} of the stuck population, and
\catThree{} under the injection split alone. The interval's upper bound,
\hiCatThreeC{}\%, lies below the registered quarter. \textbf{\Hone{} is
refuted} by the plan's rule, and by a small margin: the bar is \hOneBarGap{}
points above the interval. The margin holds under the two readings that
could move it, both in Section~\ref{sec:threats}: with the test code the
scope let through left out the upper bound is \testHOneHi{}\%, and with every
method the instrument's fault cost category 3 restored the pooled share would
be \catThreeCRestoredShare{} against \pooledCatThreeC{}; that interval was
not recomputed, and a shift of that size leaves the upper bound,
\hiCatThreeC{}\%, under the quarter. The count is far above
the minimum of \planHOneMinimum{}; it is the share that falls short. The
tiers agree (\tierHighCatThreeC{}, \tierMidCatThreeC{} and
\tierLowCatThreeC{}), and the median repository's own share,
\medianCatThreeC{} over \reposCatThreeC{} repositories with a stuck method,
is lower still, so the large repositories carry the pooled figure upward, not
down. The stuck methods are concentrated, the ten repositories with the most
holding \topTenShare{} of them, and in \reposStuckNoCatThree{} of those
\reposCatThreeC{} repositories (\reposStuckNoCatThreeShare{}) no stuck
method is rescued at all; the rescue share of the ten largest runs from
\topTenCatThreeCMin{} to \topTenCatThreeCMax{}, around the pooled figure
rather than above it. The remaining \catFourC{} stuck methods
(\catFourCShare{}) have no plan from either side;
Section~\ref{sec:explore} reads them.

\begin{table}[t]
\centering\small
\caption{The headline figures: the count over its denominator, the pooled share, the population estimate weighted by stratum with its \planLevel{} bootstrap interval over repositories within strata, and the median of each repository's own share over the repositories with a method in the denominator.}
\label{tab:headline}
\begin{tabular}{>{\raggedright\arraybackslash}p{4.7cm}rrlrr}
\toprule
Figure & Count / of & Pooled & Population (95\% CI) & Median & Repos. \\
\midrule
Stuck methods the injection split has a plan for & 21{,}350 / 38{,}555 & 55.4\% & 55.4\% (51.9 to 58.9) & 60.0\% & 470 \\
Category 3 under the injection split & 7{,}533 / 38{,}555 & 19.5\% & 19.2\% (17.5 to 21.3) & 14.6\% & 470 \\
Stuck methods the combined split has a plan for & 24{,}346 / 38{,}555 & 63.1\% & 63.2\% (60.0 to 66.4) & 66.7\% & 470 \\
Category 3 under the combined split, the rescue: \textbf{H1} & 8{,}540 / 38{,}555 & 22.1\% & 22.2\% (20.3 to 24.4) & 17.6\% & 470 \\
Improved: injection chosen and cheaper, of what SinewSlim slims: \textbf{H2} & 35{,}656 / 129{,}911 & 27.5\% & 29.0\% (24.3 to 34.6) & 21.5\% & 559 \\
Tied on the sum, settled by the tie-break & 26{,}055 / 129{,}911 & 20.1\% & 20.9\% (19.1 to 23.0) & 19.1\% & 559 \\
\bottomrule
\end{tabular}

\end{table}

\begin{table}[t]
\centering\small
\caption{The stuck methods by category, tested conventional first, under the injection split and under the combined split. Category 3 under the combined split is \Hone{}'s count.}
\label{tab:categories}
\begin{tabular}{>{\raggedright\arraybackslash}p{8.2cm}rrrr}
\toprule
Category of a stuck method & \multicolumn{2}{c}{Injection split} & \multicolumn{2}{c}{Combined split} \\
 & Count & of stuck & Count & of stuck \\
\midrule
1: slims by extraction once its jumps are ignored (conventionally fixable) & 17{,}132 & 44.4\% & 17{,}132 & 44.4\% \\
2: slims by extraction with a record & 3{,}619 & 9.4\% & 3{,}619 & 9.4\% \\
3: a plan under the OfficeFloor column only (the rescue) & 7{,}533 & 19.5\% & 8{,}540 & 22.2\% \\
4: no plan from either (needs a redesign) & 10{,}271 & 26.6\% & 9{,}264 & 24.0\% \\
\midrule
Stuck methods & 38{,}555 & & 38{,}555 & \\
\bottomrule
\end{tabular}

\end{table}

\paragraph{Why the combined split fails.} Table~\ref{tab:reasons} gives
every reason, under each split. A loop longer than the target is the reason
for \loopShareStuckNoPlanC{} of the \nStuckNoPlanC{} stuck methods the
combined split has no plan for: a loop whose body SinewSlim cannot slim
inside the function either. A single statement longer than the target,
usually a \sig{switch} of many short cases, a \sig{try} with resources or a
\sig{finally}, and a \sig{switch} that falls through account for most of the
rest. A jump crossing every cut, the obstacle that most often stops
extraction, stops the combined split on only \jumpsStuckNoPlanC{}: on jumps,
the split does what it was built to do. That is more than the
\jumpsStuckNoPlan{} the injection split alone stops at for the same reason:
\nCatThreeLost{} of them are the fault of Section~\ref{sec:threats}, and
the rest are methods the injection split stops at a loop or a statement over
the target, which the combined split may take whole as a function body, so
that the obstacle it reports moves to the jump crossing every cut outside
it.

\begin{table}[t]
\centering\small
\caption{Why the split has no plan: each reason's count among the stuck methods and among all long methods, under the injection split and under the combined split.}
\label{tab:reasons}
\fitwidth{\begin{tabular}{>{\raggedright\arraybackslash}p{5.4cm}rrrrrr}
\toprule
Reason & \multicolumn{3}{c}{Injection split} & \multicolumn{3}{c}{Combined split} \\
 & Stuck & Share & All long & Stuck & Share & All long \\
\midrule
A loop longer than the target & 8{,}352 & 48.5\% & 18{,}528 & 6{,}234 & 43.9\% & 6{,}234 \\
A single statement longer than the target & 2{,}839 & 16.5\% & 6{,}553 & 2{,}610 & 18.4\% & 2{,}612 \\
A try with resources or a finally & 2{,}500 & 14.5\% & 8{,}544 & 1{,}759 & 12.4\% & 1{,}759 \\
A switch whose cases fall through & 1{,}714 & 10.0\% & 1{,}962 & 1{,}635 & 11.5\% & 1{,}635 \\
More than 5 carried values in every plan & 1{,}623 & 9.4\% & 1{,}921 & 1{,}488 & 10.5\% & 1{,}488 \\
A jump crossing every cut & 177 & 1.0\% & 226 & 483 & 3.4\% & 497 \\
\midrule
Without a plan & 17{,}205 & & 37{,}734 & 14{,}209 & & 14{,}225 \\
\bottomrule
\end{tabular}
}
\end{table}

\paragraph{Where extraction would do: \Htwo{}.} Of the \nSlims{} long
methods SinewSlim slims, the combined split chooses injection and costs less
than SinewSlim's cost plus 1 for \nImproved{}: \ciImproved{} of the
population, a median repository share of \medianImproved{}, close to the
\priorImprovedShare{} seen on OfficeFloor's own code before the plan. Another
\nTied{} (\popTied{}) tie on the sum and go to injection by the tie-break,
and \notImprovedShare{} are left to extraction alone. Among the improved
methods the saving is a median of \savingMedian{} (quartiles \savingQone{}
and \savingQthree{}), and \savingOneShare{} of them save exactly 1, so that
their sum equals SinewSlim's cost: the extra function has paid for itself and
no more. That plus 1 is the cost of the method as one function, which a
developer who keeps the method as a method never pays, so the registered
figure is the improvement against extraction inside OfficeFloor. Against
plain extraction, the sum strictly below SinewSlim's cost itself, the share
is \nImprovedStrict{} of the \nSlims{}: \ciImprovedStrict{}, a median
repository share of \medianImprovedStrict{}. The second reading was not
registered; a reader who takes \Htwo{} as ``costs less than extraction''
should take that one.

\paragraph{Where both can split.} Where SinewSlim and the injection split
both have a plan, \nBoth{} methods, the split's worst seam carries fewer
values than SinewSlim's widest piece for \splitSmallerSeam{}, the same for
\equalSeam{} and more for \splitLargerSeam{}; the \nWorse{} methods where it
carries more are listed in the analysis, every one, and \worseByOneShare{} of
them are larger by a single value. The comparison leans towards the split by
definition, since a function's seam leaves out the method's parameters and a
piece's width counts them, and the scans cannot say by how much; and the
split gets there by cutting more, more functions than SinewSlim makes pieces
for \splitLargerParts{}.

\paragraph{Post hoc.} Nothing in the two paragraphs that follow was
registered. Each was made after the registered analysis had been read, and
each is a reading of the same scans, not a new measure.

\paragraph{Post hoc 1: where the shortfall sits.} The categories credit
injection only with what no restructuring could reach, but the combined
split has a plan for \planInCatOneShare{} of the category 1 methods and
\planInCatTwoShare{} of category 2, taking them apart as they stand, with no
guard inverted and no record written. Read over the \nNotConventional{}
stuck methods that nothing conventional fixes, categories 3 and 4 together,
category 3 is \rescueOfNotConventional{} (population
\rescueOfNotConventionalCi{}, median repository
\rescueOfNotConventionalMedian{}): Table~\ref{tab:categories} with a
different denominator, not a new measure. The registered reading stands;
this one says where the shortfall against the quarter comes from.

\paragraph{Post hoc 2: the shape of the plans injection wins.} Every plan
records its functions, each a run, a decision or a handler, and its links,
each a next, a flow or a return, so a plan can be read as a straight sequence
of functions or as branching, with or without early returns
(Table~\ref{tab:shapes}). Where extraction would do and injection is chosen
and cheaper, the plan is a sequence in \shapeImprovedAnySequence{} of the
\shapeNImproved{} methods (\shapeImprovedSequenceReturns{} of them with early
returns), and the tied plans are sequences in \shapeTiedAnySequence{}. Where
injection rescues a stuck method, the plan branches in
\shapeRescuedAnyBranching{} of the \shapeNRescued{}, and
\shapeRescuedSequenceReturns{} are sequences with early returns.
Read as a population, the long methods for which a straight sequence of
functions is the cheaper split under the registered models, the plan a
sequence and either cheaper than extraction wrapped as one function or the
only plan there is, are \ciSeqWay{} of the long methods (\nSeqWay{} in the
sample), or \ciSeqWayStrict{} against plain extraction, a saving of 2 or
more. Nearly all are methods extraction would also split; the ancillary data
has the figure with ties and for plans of any shape.

\begin{table}[t]
\centering\small
\caption{Post hoc: the shape of the combined split's plan where it uses injection, read from the plan's functions and links. A sequence has no decision function; early returns are return links from a function other than the last.}
\label{tab:shapes}
\fitwidth{\begin{tabular}{>{\raggedright\arraybackslash}p{5.2cm}rrrrr}
\toprule
Plans & Count & Sequence & Sequence, early returns & Branching & Branching, early returns \\
\midrule
Improved (injection chosen and cheaper) & 35{,}656 & 28{,}282 (79.3\%) & 6{,}092 (17.1\%) & 1{,}065 (3.0\%) & 217 (0.6\%) \\
Tied (chosen by the tie-break) & 26{,}055 & 21{,}637 (83.0\%) & 3{,}817 (14.6\%) & 560 (2.1\%) & 41 (0.2\%) \\
Rescued (category 3, combined split) & 8{,}540 & 2{,}202 (25.8\%) & 854 (10.0\%) & 3{,}846 (45.0\%) & 1{,}638 (19.2\%) \\
\bottomrule
\end{tabular}
}
\end{table}

\section{Discussion}

The author expected function injection to take apart a good share of what
extraction cannot, and set a quarter as the bar. On \nUsable{} repositories
of code written before AI assistants were in general use, the registered
reading is \popCatThreeC{}, and the interval excludes the quarter, by
\hOneBarGap{} points. The expectation was right in direction against the one
repository seen beforehand (\priorCatThreeShare{} on the framework's own
code) and wrong in size against the bar. A reader sceptical of OfficeFloor
should take the figure as it stands: function injection is not a general fix
for stuck methods.

The post hoc reading says where the shortfall sits. Nearly half of the stuck
methods (\catOneShare{}) are methods SinewSlim would slim if their jumps were
ignored: the guard-clause and early-return methods, of which \sig{describe}
in Figure~\ref{fig:examples} is one. The plan put them in category 1,
conventionally fixable, which was the reading least favourable to injection
and the right one to register, and it stands. What that category assumes is
that ignoring a jump stands for a refactoring a developer would make. The
nearest ones are inverting the guards, threading a flag through nested
\sig{if}s, or extracting the run with a flag the caller tests and returns on,
as an IDE does for one early return; the study prices none of them, and the
first two many would call the worse design. The combined split handles the
jump as it stands, and takes apart three in four of those methods without
restructuring them. Of the stuck methods that nothing conventional fixes at
all, it takes apart about half. That is the clearest pattern in the data:
single-exit extraction, as SinewSlim models it, cannot deal with an early
return, and a flow can. What is left after that, \catFourCShare{} of the
stuck methods, has no plan from either side, and Section~\ref{sec:explore}
reads what it holds.

The shape of the plans separates the two results by mechanism. Where
injection is chosen over extraction, the method is almost always a sequence:
stages in order, each consuming what the one before produced, sometimes
leaving early. The reading of Equations~\eqref{eq:extraction}
and~\eqref{eq:injection} in Section~\ref{sec:background} says where such a
sequence is cheaper than a method lifted into pieces, and it is not at every
hand-off: where several values cross one boundary, where a method's
parameters are read past the first cut, and where the last piece hands back
the method's own result; \sig{build} in Figure~\ref{fig:examples} shows
the second. The saving is small because most
boundaries carry one or two values: a median of \savingMedian{}, and
\savingOneShare{} of the improved methods save exactly 1, the wrapper
function's own cost. The claim the data supports is that the long methods
people write are often pipelines, and that under the registered cost models
a pipeline whose stages hand on more than one value, or read the method's
parameters throughout, is cheaper written as a sequence of functions than
contained in a method and lifted into pieces. The scans do not record which
source each saving comes from, so their shares are not known.
The rescue has the other mechanism: the methods only injection takes apart
are mostly branching, an \sig{if} or \sig{switch} whose arms become flows,
\shapeRescuedReturnsOfBranching{} of them with early exits, which is where
extraction's single-exit piece fails outright.

As a population, the methods for which a sequence of functions is the
cheaper split are \popSeqWay{} of the long methods, or \popSeqWayStrict{}
against plain extraction. At the margin, a saving of 1, the two designs cost
the same under the registered models, and what the sequence adds is not
priced. Under injection the caller becomes configuration: the order of the
functions and what each receives are declared, rather than read from
statement order and the local variables that carry one stage's product to
the next. Whether a declared flow is easier to follow than the method it
replaces is a question about reading, which nothing here measures; the
\nSeqWay{} methods are where such a study would look. Every method here was
written by people, before AI coding assistants were in general use; whether
code written with assistants is stuck as often, and in the same ways, is a
study for which the figures here are the baseline.

\section{What the redesign list holds}\label{sec:explore}

Nothing in this section was registered or bears on a registered figure. It
was read after the results, to decide what the tools should do with the
methods they stop at, by a separate step whose harness, scripts and results
are in the study's repository; Section~\ref{sec:next} says what follows
from it. The
\catFourC{} category 4 methods, \catFourCShare{} of the stuck methods and
\exRateOverall{} of the long methods, are those SinewSlim cannot slim and
the combined split has no plan for. A harness re-read \exReread{} of them
from source with a parser of its own and classed each by the innermost
obstacle; at the scan's limit of five it gives none a combined plan, so it
reproduces the scan, and it refines the instrument's reason into the classes
of Table~\ref{tab:explore-classes}.

\begin{table}[t]
\centering\small
\caption{Exploratory: the category 4 methods by class, the largest classes grouped by family. \emph{Top ten} is the share of the class held by the ten repositories with the most of it, beside the same repositories' share of all long methods (what size alone predicts); \emph{dispersion} is 1 when the class is spread as the long methods are and higher when clustered; \emph{cognitive per line} is the class's median cognitive complexity \cite{campbell2018} per code line.}
\label{tab:explore-classes}
\fitwidth{\begin{tabular}{>{\raggedright\arraybackslash}p{6.6cm}rrrrrrr}
\toprule
Class & Methods & Share & Repos. & Top ten & Size predicts & Dispersion & Cognitive per line \\
\midrule
\textbf{Tables and literals} & \textbf{3{,}030} & \textbf{32.7\%} & & & & & \\
\quad A switch of short cases, too many together & 1{,}028 & 11.1\% & 142 & 45\% & 32\% & 4.2 & 0.21 \\
\quad A switch that is a flat table (every case 3 lines or fewer) & 346 & 3.7\% & 97 & 47\% & 30\% & 1.8 & 0.12 \\
\quad An array initialiser & 416 & 4.5\% & 44 & 87\% & 18\% & 11.5 & 0.00 \\
\quad A long argument list & 390 & 4.2\% & 58 & 65\% & 13\% & 9.9 & 0.00 \\
\quad A builder chain & 388 & 4.2\% & 76 & 59\% & 15\% & 5.7 & 0.00 \\
\quad A string concatenation & 260 & 2.8\% & 65 & 64\% & 20\% & 11.4 & 0.00 \\
\quad A constructor that only delegates & 81 & 0.9\% & 19 & 86\% & 22\% & 1.7 & 0.00 \\
\quad A body of one statement & 121 & 1.3\% & 35 & 60\% & 8\% & 2.3 & 0.00 \\
\textbf{Dense logic} & \textbf{2{,}341} & \textbf{25.3\%} & & & & & \\
\quad A break or continue inside a loop & 387 & 4.2\% & 101 & 50\% & 30\% & 1.5 & 0.57 \\
\quad A return inside a loop & 204 & 2.2\% & 75 & 54\% & 30\% & 0.9 & 0.57 \\
\quad A return and a break or continue inside a loop & 157 & 1.7\% & 56 & 56\% & 28\% & 1.7 & 0.63 \\
\quad A return inside a try & 172 & 1.9\% & 49 & 62\% & 31\% & 1.5 & 0.41 \\
\quad An if that passes values on & 170 & 1.8\% & 57 & 56\% & 30\% & 1.0 & 0.53 \\
\quad An if that passes values on and jumps & 97 & 1.0\% & 43 & 62\% & 27\% & 1.3 & 0.59 \\
\quad Too many values at every cut, even at a limit of 10 & 1{,}154 & 12.5\% & 159 & 46\% & 34\% & 3.0 & 0.46 \\
\textbf{Dispatch and nesting} & \textbf{773} & \textbf{8.3\%} & & & & & \\
\quad An else-if chain (dispatch) & 375 & 4.0\% & 107 & 44\% & 32\% & 3.9 & 0.44 \\
\quad A switch with a case over the cap & 154 & 1.7\% & 53 & 60\% & 28\% & 1.4 & 0.41 \\
\quad A nested try over the cap & 129 & 1.4\% & 26 & 88\% & 24\% & 1.8 & 0.08 \\
\quad A nested loop over the cap & 115 & 1.2\% & 43 & 61\% & 15\% & 2.4 & 0.47 \\
\textbf{Data within reach of a higher limit} & \textbf{2{,}266} & \textbf{24.5\%} & & & & & \\
\quad A plan at a limit of 6 to 10 & 1{,}132 & 12.2\% & & & & & \\
\quad Only the blocking body slims, at a limit of 6 to 10 & 1{,}134 & 12.2\% & & & & & \\
All other classes & 854 & 9.2\% & & & & & \\
\midrule
Category 4 & 9{,}264 & & 331 & 39\% & 38\% & 7.8 & 0.36 \\
\bottomrule
\end{tabular}
}
\end{table}

\paragraph{Three causes.} \exIndivisibleShare{} of the list (\exIndivisible{}
methods, in \exIndivisibleRepos{} repositories) is \emph{indivisible}: one
statement over the cap that nothing inside can move. \exDataShare{}
(\exData{}) is \emph{data}: every cut needs more than five values.
\exControlShare{} (\exControl{}) is \emph{control}: a return, break or
continue crosses every cut. The instrument's reason \emph{jumps} also covered
\exRegrouped{} methods where nothing can be cut at all, a body of one
statement, a constructor that only delegates, or returns that belong to a
lambda; the harness moves these to indivisible, and no conclusion below
depends on the move.

\paragraph{Tables and literals.} A third of the list, \exTables{} methods
(\exTablesShare{}), is tables and literals: \sig{switch} statements of short
cases (\exSwitchShort{}) or that are flat tables (\exSwitchFlat{}), array
initialisers, long argument lists, builder chains, string concatenations,
constructors that only delegate, and bodies of a single expression. They are
long for their data or their options, not their logic: their cognitive
complexity is zero or near it, \exCogPerLineFlat{} and \exCogPerLineShort{}
per line for the two kinds of switch against \exCogPerLineCatFour{} for the
list as a whole. Of the \exSwitches{} switches in the list
(\exSwitchesShare{}), \exSwitchOptions{} are long only because of how many
options they have; only the \exSwitchCaseOver{} with a case over the cap
have a body to split. Read as methods of their own, the blocking statements
of the literal classes take data in and call nothing internal
(\exLitCatchmentArray{} of the array initialisers, \exLitCatchmentConcat{}
of the string concatenations), where those of the logic classes hand on to
other internal code (\exLogicJunctionCase{} of the long switch cases,
\exLogicJunctionTry{} of the nested tries): the logic has seams, and is
blocked by jumps and carried values, not by a lack of them. Arguably the
tables are not long methods in the sense the trigger intends.

\paragraph{Dense logic.} The rest is where a split would help a reader, and
the harness says why it fails. In the control group,
\exControlContainedShare{} (\exControlContained{} of \exControl{}) have the
blocking jump inside a loop, \sig{try} or \sig{switch}: \exControlLoop{} in a
loop, \exControlTry{} in a try, \exControlSwitch{} in a switch. The injection
split cuts a sequence of steps, and a step whose body breaks out of its
enclosing loop or returns from the method cannot become a function, so the
loop stays whole. Only \exControlTop{} have the jump at the method's top
level, \exControlTopReturn{} of them long branching methods with early
returns. These are the densest methods in the study, \exCogPerLineLoopLo{}
to \exCogPerLineLoopHi{} cognitive complexity per line in the loop classes.
Of the \exIfs{} methods blocked by an \sig{if} (\exIfsShare{}),
\exIfDispatch{} are else-if chains that dispatch; of the \exIfNonDispatch{}
others, \exIfPassOrJump{} (\exIfPassOrJumpShare{}) pass values on from their
branches, jump, or both, and they are long and dense (median
\exIfDenseLengthLo{} to \exIfDenseLengthHi{} lines, \exIfDenseCplLo{} to
\exIfDenseCplHi{} per line). A branchy leaf, an \sig{if} that passes nothing
out and jumps nowhere, is the rarest kind (\exIfLeaf{}, \exIfLeafShare{} of
the list). What blocks an \sig{if} is its outputs, not its branching.

\paragraph{The limit of five.} The data group was re-run with the limit
raised (Table~\ref{tab:explore-limits}). Six gives the combined split a plan
for \exLimitSixN{} of the list (\exLimitSix{}), the largest single gain; ten
for \exLimitTenN{} (\exLimitTen{}). Within the data group,
\exDataPlanShare{} (\exDataPlan{}) gain a plan at some limit up to ten; for
\exDataBodyShare{} (\exDataBody{}) only the blocking statement slims as a
body of its own and the method as a whole still has no plan; and
\exDataNoneShare{} (\exDataNone{}) have no plan even at ten. The last are
the most complex of the data classes (\exDataNoneCogPerLine{} per line,
median \exDataNoneLength{} lines) and need a different split, not a looser
limit. SinewSlim alone gains at about the same rate (\exLimitSixConventionalN{}
at six), so a higher limit moves categories 1 to 3 as well as 4, which is
why Section~\ref{sec:next} proposes sweeping it over the whole population
rather than raising it here.

\begin{table}[t]
\centering\small
\caption{Exploratory: of the category 4 methods, how many have a plan when the limit of five is raised, under the combined split, under SinewSlim alone (its parameters a piece) and under the injection split alone (its carried values a function).}
\label{tab:explore-limits}
\begin{tabular}{lrrrrrr}
\toprule
Limit & \multicolumn{2}{c}{Combined split has a plan} & \multicolumn{2}{c}{SinewSlim slims} & \multicolumn{2}{c}{Injection split has a plan} \\
 & Methods & Share & Methods & Share & Methods & Share \\
\midrule
5 & 0 & 0.0\% & 0 & 0.0\% & 189 & 2.0\% \\
6 & 398 & 4.3\% & 328 & 3.5\% & 442 & 4.8\% \\
7 & 690 & 7.4\% & 635 & 6.9\% & 648 & 7.0\% \\
8 & 889 & 9.6\% & 900 & 9.7\% & 790 & 8.5\% \\
10 & 1{,}132 & 12.2\% & 1{,}277 & 13.8\% & 1{,}015 & 11.0\% \\
\bottomrule
\end{tabular}

\end{table}

\paragraph{Complexity.} At the same length, category 4 is no more complex
than the methods injection rescues. Standardised to category 4's length
distribution, its mean cognitive complexity is \exCogCatFour{}, against
\exCogRescued{} for category 3, \exCogConventional{} for categories 1 and 2,
and \exCogSlimmable{} for the methods SinewSlim slims; per line,
\exCogPerLineCatFour{} against \exCogPerLineRescued{} and
\exCogPerLineConventional{}. What keeps a method in the list is its shape,
not its complexity, and the list holds two families: the tables, with no
complexity to speak of, and the dense logic, at about
\exDataNoneCogPerLine{} to \exCogPerLineLoopHi{} per line.

\paragraph{Where the list is.} The list is in \exRepos{} repositories, so
it is widespread, and it is also clustered beyond what repository size
explains. Its dispersion index, the chi-square of each repository's count
against the count its long methods predict, over the repositories with a
long method less one, is \exDispersion{}, where 1 would mean spread as the
long methods are; the ten repositories with the most hold
\exTopTenShare{}, close to the \exTopTenSizeShare{} their share of long
methods predicts, so the clustering is not only the largest few; and
\exHotRepos{} repositories whose standardised residual against the pooled
rate exceeds 3.9, about the Bonferroni bound for \nUsable{} tests at the
plan's level, hold \exHotShare{} of the list from \exHotLongShare{} of the
long methods. The classes split by how they are spread
(Table~\ref{tab:explore-classes}). The literal classes are concentrated:
the ten repositories with the most array initialisers hold
\exArrayTopTen{} of them where size predicts \exArrayTopTenSize{}, a
dispersion of \exArrayDispersion{}, and the long argument lists, string
concatenations and nested tries (\exNestedTryTopTen{} in ten repositories)
are alike, which looks like generated code or a house style of building
data in code. The logic classes are spread as the long methods are, flat
switches at \exFlatDispersion{} and ifs that pass values on at
\exIfPassDispersion{}, which looks like ordinary ways of writing code. A
reading by domain, coded by keyword from each repository's description and
topics, is in the ancillary data; its intervals are wide. Two repositories,
\exBudgetRepos{}, ran out of the call-resolution budget the statement
readings use, so their internal calls are undercounted; the complexity
figures do not depend on call resolution.

\section{What follows}\label{sec:next}

The redesign list is where both tools stop, and Section~\ref{sec:explore}
says it is not one thing. Three studies follow from it, each to be
registered as this one was, by a dated commit of its plan, seed and reading
rule in a public repository before its sample is drawn.

\paragraph{Study 1: the classes as rules.} The exploratory classes move
into the instrument as rules with fixed thresholds, so that a method the
split cannot plan is reported with its class and the advice that fits it.
Two hypotheses: that a fresh sample drawn the same way gives the same class
shares, within the intervals here; and that reviewers shown a method with
its class accept it, the tables first, since if reviewers do not call a
table a long method the trigger itself changes.

\paragraph{Study 2: a refinement per class, and what is left.} Each class
gets the change that reaches it, tested on the methods already listed.
Table~\ref{tab:rules} gives the classes of Table~\ref{tab:explore-classes}
with the rule the harness finds each by and what the study does with it:
named exits for a jump inside a loop, \sig{try} or \sig{switch}
(\exControlContained{} methods); the branches of an \sig{if} that passes
values on as functions with the value as the flow's argument
(\exIfPasses{}); and for data coupling (\exData{}) the limit of five swept
from three to ten over the whole population, since
Section~\ref{sec:explore} shows a higher limit moves categories 1 to 3 as
well as 4. Each refinement is read by how many of its class become
slimmable and by how cognitive complexity per line moves, since a rewrite
that slims by growing denser has not helped. What is left when every class
has had its refinement is the proposal for a definition of a god method
that needs no human judgement: a method over the cap that no split at the
chosen limit and no rule-following rewrite takes apart. The hypothesis is
that the residual is under one in a hundred long methods, and dense logic
rather than data.

\begin{table}[tbp]
\centering\small
\caption{The classes of the redesign list, the rule the exploratory harness finds each by, and what the second study does with it. The root cause is the instrument's reason: a statement over the cap that nothing inside can move (\emph{indivisible}), more than five carried values in every plan (\emph{data}), or a jump crossing every cut (\emph{control}); a control method whose jump crosses no cut moves to indivisible. The rules are applied in the order listed within each family, to the innermost statement over the cap; a cut is a boundary between statements.}
\label{tab:rules}
\fitwidth{\begin{tabular}{>{\raggedright\arraybackslash}p{3.3cm}>{\raggedright\arraybackslash}p{7.0cm}>{\raggedright\arraybackslash}p{4.9cm}}
\toprule
Class & Rule & Refinement \\
\midrule
\multicolumn{3}{l}{\emph{Tables and literals} (indivisible)} \\
A constructor that only delegates & The body is a single \sig{super(\ldots)} or \sig{this(\ldots)} call & \multirow{8}{4.9cm}{Out of the trigger, by the first study: a table to leave in place or move to configuration} \\
A body of one statement & One top-level statement that holds no others: a return or an expression & \\
A switch that is a flat table & A \sig{switch} with every case three lines or fewer & \\
A switch of short cases & A \sig{switch} with no case over the cap and a case over three lines & \\
An array initialiser & An expression whose array initialiser spans at least half its lines & \\
A builder chain & An expression with a chain of four or more calls & \\
A string concatenation & An expression concatenating five or more string literals & \\
A long argument list & An expression with a call of eight or more arguments & \\
\midrule
\multicolumn{3}{l}{\emph{Dense logic} (control and data)} \\
A break or continue inside a loop; a return inside a loop; both & The blocking jump leaves a loop, or targets a loop or switch nested in it; by the kind of jump & Named exits: the jump becomes the function's named exit, wired in configuration \\
A return inside a try & The same, with a \sig{try} & The same \\
An if that passes values on & An \sig{if} with fewer than two \sig{else if} arms, values set in its branches read after it, and no jump leaving it & The branches become functions, the value the flow's argument \\
An if that passes values on and jumps & The same, with a jump leaving it & The same \\
Too many values at every cut, even at a limit of ten & The instrument's reason is carried values, and the combined split has no plan at any limit up to ten & A different split, not a looser limit; what the sweep leaves \\
\midrule
\multicolumn{3}{l}{\emph{Dispatch and nesting} (indivisible)} \\
An else-if chain & An \sig{if} with two or more \sig{else if} arms & Classed and reported by the first study; no refinement \\
A switch with a case over the cap & A \sig{switch} with a case over the cap & The case extracted as a method of its own \\
A nested try or loop over the cap & The innermost statement over the cap is a \sig{try} or a loop whose body nothing can move & Classed and reported by the first study; no refinement \\
\midrule
\multicolumn{3}{l}{\emph{Data within reach of a higher limit} (data)} \\
A plan at a limit of 6 to 10 & The combined split has a plan at the first limit from six to ten & The limit swept from three to ten over the whole population \\
Only the blocking body slims at 6 to 10 & No plan at ten, but SinewSlim slims the blocking statement as a body of its own at some limit from six to ten & The same \\
\bottomrule
\end{tabular}}
\end{table}

\paragraph{Study 3: the measures on code written with assistants.} A
sample drawn by the same design from repositories whose code was written
after assistants were in general use, scanned with the refined instrument,
and its long share, stuck share, class shares and residual compared with
the figures here, which are its baseline. The claim is bounded to
decomposition: the measures say nothing about how methods couple to one
another or whether the code will change well.

\section{Threats to validity}\label{sec:threats}

Stuck is relative to the extraction SinewSlim searches: contiguous pieces
with one exit, one output and at most five parameters. An IDE's Extract
Method takes a run with an early return by handing back a flag or an
\sig{Optional} that the caller tests and returns on, and a developer can
write such a piece by hand; SinewSlim does not search that form, so a method
stuck here may come apart under an extraction it does not price. Category 1,
\catOneShare{} of the stuck methods, is where this matters, and the count of
what nothing conventional fixes is a count under one tool's rules. The limits
every count rests on, five parameters a piece, five carried values a
function, a trigger of 26 and a target of 25, are fixed points: the trigger
is the author's own earlier finding, and Section~\ref{sec:explore} shows a
limit of six gives a plan to \exLimitSix{} of the redesign list, so each
would move the categories; Section~\ref{sec:next} proposes sweeping the
limit for that reason. The split is a projection: none of this code was
written for function injection, so a plan is what the split would cost, not
what anyone paid, and the migration is not analysed here. The injection
split's limits decide much: loops and lambda bodies are never cut between
functions, a choice made before the plan for the framework's own reasons,
and Section~\ref{sec:explore} shows it is where most of the control group
sits.

The baseline \Htwo{} is read against is chosen by a different rule from the
plan it is set beside. SinewSlim takes the segmentation with the smallest
worst seam first and the smallest cost last, and the combined split takes
the plan with the smallest sum; where a segmentation of wider seam and lower
cost than SinewSlim's exists, it is not the baseline, so \Htwo{} leans
towards injection by the width of that gap, which the scans do not record.
The combined split's tie-break favours injection, so ties are reported apart
and never counted, and its one-to-one weight between the two totals is a
choice; a reader who weighs a flow dearer than a parameter would see fewer
improved methods. \Htwo{}'s baseline of SinewSlim's cost plus 1 includes a
function a developer outside OfficeFloor never pays for, so the reading
against SinewSlim's cost itself is reported beside it. The cost models price
what crosses a boundary and nothing about how the result is read, so
whatever a declared flow adds to understanding, or takes away, is outside
every figure here.

The instrument has one fault in how it builds a function body's facts, found
after the registered analysis and reported to its project. The combined
split has no plan for \nExceptions{} of the \nSlims{} methods SinewSlim slims
(\exceptionsShare{}), which the plan says cannot happen by construction, and
no plan for \nCatThreeLost{} of the \catThree{} methods the injection split
has a plan for, all with the reason \emph{jumps}; both lists are in the
ancillary data, and both count against injection. The most the second could
add to \Hone{} is every lost method restored to category 3,
\catThreeCRestoredShare{} pooled, under the quarter. The split also charges
a flow for an early return in a function that otherwise falls off the
method's end, where the flow goes where falling off the end goes; it can
only overstate injection's cost. The scope's globs let through test code in
directories named otherwise (\sig{\testTopSegment},
\sig{\testSecondSegment}): \testDroppedLong{} long methods
(\testDroppedLongShare{}) in \testRepos{} repositories. With every method
under a directory whose name contains \emph{test} left out, the rescue is
\testHOneCi{} and the improvement \testHTwoCi{}, and neither reading
changes. Generated code predates AI assistants and is in the sample wherever
its directory is not named \sig{generated}; the concentrated literal classes
of Section~\ref{sec:explore} are where it shows, and the longest method in
the sample, \longestMethodLines{} lines, is a table in
\sig{\longestMethodRepo}. Stars measure attention, not craft, and a
repository's tier is its popularity now, not at the as-of date. Assistants
older than Copilot existed before the as-of date, so the sample is code
written before AI code generation was in general use, not before any of it.

The registration is a commit in a repository the author controls, pushed to
GitHub on 1 November 2026; the evidence of its date is GitHub's record of
that push and of the commits that follow it, not an independent registry's
timestamp. A reader who wants more than that has the plan as registered,
deposited with the data, and the scripts, from which every step from the
draw to this paper's numbers is repeated without the author. The exploratory reading
is one pass over one list by its author, its class thresholds were chosen by
eye to sort that list and are not yet validated, which is the first
proposed study's job. The studies of Section~\ref{sec:next} are proposals,
not findings. The study is Java only, and
the author created the framework and both products: this is why the sample
was drawn by a separate tool with a seed fixed in the plan, why every cost
is the products' and fixed before the plan, why the two totals are never
compared, and why every reason the split fails and every method where it
does worse is reported.

\section{Conclusion}

On \nUsable{} open-source Java repositories as they stood before AI coding
assistants were in general use, \stuckShare{} of the long methods cannot be
slimmed by single-exit Extract Method as SinewSlim models it. The registered
hypothesis that function injection takes apart at least a quarter of them is
refuted: it takes apart \ciCatThreeC{}, the bar \hOneBarGap{} points above
the interval. A reading not registered says where the shortfall sits:
nearly half of the stuck methods are early-return methods the plan counted
as conventionally fixable, and the split takes three in four of those apart
as they stand. Where extraction would do and injection is chosen instead,
the method is almost always a pipeline, cheaper under the registered cost
models as a sequence of functions because a value crossing a boundary is
priced once as a declared value and twice as an output and an input, and a
parameter read throughout is priced by extraction alone. The quarter of the
stuck methods that no split takes apart is, on an exploratory reading, a
third tables and literals that are long without being complex, and
otherwise logic blocked by a jump inside a loop or \sig{try}, by an
\sig{if} whose branches pass values on, or by more values than any cut may
carry; three studies are proposed to turn that reading into rules, a
refinement per class, and a baseline against code written with assistants.
The author expected more of the rescue, and the plan was written so that
the data could say so.

\paragraph{Data availability.} Every number here is produced by a script
from the committed analysis, and every table is a generated fragment; the
ancillary data holds the registered plan with its amendments, the sample with
its derivation, the run log, the registered analysis with every
per-repository figure and every method where the split's worst seam is
larger, the post hoc readings with the methods they list, the exploratory
reading of category 4 with its tables and domain coding, and the two worked
examples with their scan. \ifdefined\plandoi The plan as it stood at the
registration commit is deposited at \texttt{doi:\plandoi}. \fi
\ifdefined\scansdoi The raw scans, one file per repository, are deposited at
\texttt{doi:\scansdoi}, so that a replication does not depend on the
\nUsable{} repositories remaining on GitHub at their pinned commits; they
\else The raw scans \fi
are reproduced by the study's scripts from the pinned sample and the pinned
jar, so a replication needs no cooperation from the author.

\section*{Competing interests and funding}

The author created OfficeFloor, SinewSlim and OfficeFloorSlim, and has a
commercial interest in OfficeFloor's adoption. No external funding was
received.

\end{document}